# Selection rules in the excitation of the divacancy and the nitrogen-vacancy pair in 4H- and 6H-SiC

Danial Shafizadeh[1], Joel Davidsson[1], Takeshi Ohshima[2,3], Igor A. Abrikosov[1], Nguyen T. Son[1], and Ivan G. Ivanov[1,*]

[1)] Department of Physics, Chemistry and Biology, Linköping University, SE-58183, Linköping, Sweden

[2)] National Institutes for Quantum Science and Technology, 1233 Watanuki, Takasaki, Gunma 370-1292, Japan

[3)] Department of Materials Science, Tohoku University, 6-6-02 Aramaki-Aza, Aoba-ku, Sendai 980-8579, Japan

*Author to whom correspondence should be addressed: ivaiv28@liu.se

**Abstract:** In this study, we address the selection rules with respect to the polarization of the optical excitation of two colour centres in 4H-SiC and 6H-SiC with potential for applications in quantum technology, the divacancy and the nitrogen-vacancy pair. We show that the photoluminescence (PL) of the axial configurations of higher symmetry ($C_{3v}$) than the basal ones ($C_{1h}$) can be cancelled using any excitation (resonant or non-resonant) with polarization parallel to the crystal axis ($E_L \| c$). The polarization selection rules are determined using group-theoretical analysis and simple physical arguments showing that phonon-assisted absorption with $E_L \| c$ is prohibited despite being formally allowed by group theory. A comparison with the selection rules for the silicon vacancy, another defect with $C_{3v}$ symmetry, is also carried out. Using the selection rules, we demonstrate selective excitation of only one basal divacancy configuration in 4H-SiC, the P3 line and discuss the higher contrast and increased Debye-Waller factor in the selectively excited spectrum.

## I. INTRODUCTION

Since the last few decades, the development of solid-state quantum bits (qubits) based on colour centres in wide-bandgap semiconductors, has been led by the negative nitrogen-vacancy (NV⁻) centres in diamond, showing impressive progress in applications to nano-scale sensing and quantum communications [1-2]. However, no individual colour centre or material platform is ideal for all different applications. With time, more materials with specific advantages attract attention and join the race [3,4]. Among these, silicon carbide (SiC) is a promising one due to that it hosts various colour centres emitting light near and at telecom wavelengths with excellent optical and spin properties and, more importantly, for large-scale applications, it is the only wide-bandgap semiconductor that has industrial wafer-scale productions with well-controlled doping, mature CMOS technology, and established nanofabrication techniques [3-6].

Among various colour centres in SiC, the Si vacancy ($V_{Si}$) [7] and the neutral divacancy ($V_C V_{Si}^0$ or simply VV) [8], i.e., an uncharged complex of a Si vacancy ($V_{Si}$) and a nearest C vacancy ($V_C$), are most studied. The divacancy spins have long coherence times, ranging from milliseconds in natural SiC [9] to several seconds in isotope-purified materials [10], and can be optically controlled at room temperature or even up to 550 K [11], making it suitable for broad temperature-range quantum sensing. Having emissions in the near-infrared spectral region around 1100 nm with a high-fidelity spin-to-photon interface [12] and a good control of single nuclear spins [13], the neutral divacancy is also promising for quantum applications.

An analogous defect to the NV$^-$ centre in diamond is the negative N-vacancy centre in SiC, i.e., the negatively charged complex between a N shallow donor at a C lattice site and a nearest Si vacancy (N$_C$V$_{Si}^-$, denoted hereafter just NV) [14]. The centre has an electronic structure analogical to that of the NV$^-$ centre in diamond but emits light near the O-band of telecom wavelengths (~1176-1243 nm) [15] which is more favourable for long-distance quantum communications.

The divacancy (VV) and NV centres have four and six different configurations in 4H- and 6H-SiC, respectively. Each has two axial configurations with C$_{3v}$ symmetry, corresponding to the pairs with both constituents occupying the hexagonal sites (*hh*) or quasi-cubic sites (*kk*), and two basal configurations (*hk*) and (*kh*) with C$_{1h}$ symmetry. Correspondingly, there are four ZPLs in 4H-SiC: PL1 (*hh*), PL2 (*kk*), PL3 (*hk*), PL4 (*kh*) for VV [8] and NV1 (*hk*) NV2 (*kk*), NV3 (*hh*), NV4 (*kh*) for NV$^-$ [16], and six ZPLs for both the VV and NV in 6H-SiC [17,18].

In non-resonant excitation experiments, the phonon-assisted absorption process is also involved. Since this process is also subject to selection rules, some defect configurations with a certain symmetry may not be excited. This is reflected in reported PL spectra, which often show ZPLs with very different intensities depending on the polarization of the excited laser [15,18,19]. Weak ZPLs excited with inappropriate polarization may appear in PL spectra of ensembles but can easily be missed in a scan for single emitters. Non-resonant excitation is a common way for activating and stabilizing the bright charge state of colour centres and is needed for obtaining their PL spectra. Identification of the selection rules for excitation including non-resonant excitation of ZPLs of defect centres with different symmetries is important for optimizing the PL or ODMR detection of individual ZPLs. In this work, we investigate the selection rules in non-resonant excitation for the axial (C$_{3v}$) configurations of VV and NV centres in 4H- and 6H-SiC and compare with the selection rules for the silicon vacancy which also has C$_{3v}$ symmetry.

## II. EXPERIMENTAL

The samples used for the studies on the NV pairs are bulk N-doped (in the range of $10^{17}$ cm$^{-3}$) 4H- and 6H-SiC, irradiated and annealed to form NV pairs. The samples containing divacancies are either as-grown bulk high-purity semi-insulating (HPSI) materials, or electron irradiated to fluences in the $10^{17} - 10^{18}$ cm$^{-3}$ range HPSI bulk 4H- and 6H-SiC substrates, annealed to about 800 °C for 20 – 30 min to create divacancies. PL measurements are carried out using a Jobin Yvon HR460 monochromator equipped with 1200 and 300 g/mm gratings, an InGaAs multi-channel detector. The excitation laser is a tuneable Ti-sapphire laser. Most spectra presented here are obtained with laser at ~ 930 nm unless stated otherwise. The samples are mounted in a variable temperature closed cycle cryostat and cooled down to 3.8 K. The resolution of ~4 Å with the 300 g/mm grating is sufficient for most of the measurements, albeit a higher resolution of ~1 Å has been needed using the 1200 g/mm for some closely spaced lines in 6H-SiC. A half-wave plate is placed in the path of the laser for rotating the polarization, and a polarizer is used in the emission path. The laser line is filtered using a 1000 nm long-pass filter.

## III. THEORETICAL BACKGROUND

The polarization selection rules for the zero-phonon line (ZPL) emission, as derived from the symmetries of the ground and excited states and group-theoretical analysis, have been discussed for many colour centres in previous works. For the divacancy and the $V_{Si}$ such consideration can be found in Refs. [20] and [21], respectively. From a group-theoretical point of view, the selection rules for the divacancy and the NV centre are the same because their ground and excited states have the same symmetry. Furthermore, the same selection rules apply for resonant excitation in the ZPLs since the matrix elements for absorption into and emission in a ZPL are the same.

However, the selection rules for non-resonant excitation of a certain defect are rarely considered, and when there is such consideration, it is usually limited to experimental observations only (e.g., [22,23]). Non-resonant excitation refers to excitation at higher energy than the ZPL into the phonon sideband (PSB) of the defect. Hence, it is phonon assisted with creation of a phonon in the absorption process, and one may assume that absorption will always be allowed if this phonon has a suitable symmetry. In this work, we refute this notion by showing that the phonon-assisted absorption into the excited state of some defects is also subject to strict selection rules. This may lead to configurations of the polarization of the exciting laser light with respect to the c-axis for which the absorption from the ground to the excited state of the defect is strictly prohibited in the case of phonon assisted transitions even though a formal group-theoretical analysis classifies it as allowed.

We examine here two different defects in the hexagonal 4H- and 6H-SiC polytypes: the neutral divacancy ($VV^0$, denoted shortly VV) and the negative nitrogen-vacancy pair ($NV^-$, or just NV) and compare with the silicon vacancy $V_{Si}$ which has been treated elsewhere [17, 23,24]. The presence of inequivalent lattice sites for Si and C, two in 4H-SiC and three in 6H-SiC, yields several defect centres (configurations) for the same type of defect. Thus, the divacancy and the NV pair in 4H- (6H-SiC) have four (six) inequivalent configurations, respectively. The silicon vacancy occupies either hexagonal or cubic Si sites, yielding two inequivalent configurations in 4H-SiC and three in 6H-SiC. All abovementioned defects have distinct ZPLs, and in some cases, the PL centres have been associated with specific configurations.

The configurations of VV and NV are further divided into axial and basal. The two (three) axial configurations possess $C_{3v}$ symmetry, in 4H- (6H-) SiC, respectively. The remaining two (three) configurations in 4H- (6H-) SiC have lower symmetry, $C_{1h}$. Our study focuses on the high symmetry ($C_{3v}$) axial configurations. The term "axial" reflects the fact that the orientation of the $C_3$ axis coincides with the crystal axis (c-axis), along which the two constituents of the binary defects are aligned. We will show that the orientation of the exciting photon's polarization with respect to the c-axis determines the probability for photon absorption, with this probability vanishing for polarization parallel to the c-axis (E∥c) in the case of both the divacancy and the NV pair in their axial configurations. However, the obtained selection rules are likely to be approximately valid also for the basal configurations with lower $C_{1h}$ symmetry [25,26], but with different orientation of the polarization direction for which the absorption nearly vanishes. We underline that we consider here usual non-resonant excitation with photon energies above the energy of the corresponding ZPL. The selection rules for resonant excitation into the ZPL are simpler and follow the selection rules for the photons emitted in the ZPL.

More details on the group-theoretical derivation of the selection rules are given in the Supplementary Material (SM file [33]), covering both cases of resonant and non-resonant excitation. The results from the analysis are summarized in Table I for the three defects considered here, $V_{Si}^-$, and the axial configurations of $V_CV_{Si}^0$, and $NV^-$.

Table I. Selection rules for the divacancy and the NV pair in their axial configurations, and the silicon vacancy $V_{Si}$ analysed within the single $C_{3v}$ group. A and F denote the allowed and forbidden transitions, respectively. The transition with E∥c assisted by phonons of E symmetry (denoted by $A^*$) is formally allowed by group theory but prohibited due to physical reasons discussed in text.

| NV pair or the divacancy (axial, $C_{3v}$ symmetry): $^3A_2 \leftrightarrow\, ^3E$ transition | | | | |
|---|---|---|---|---|
| Polarization | Resonant excitation or ZPL | Phonon assisted (non-resonant) involving phonon of symmetry | | |
| | | $A_1$ | $A_2$ | E |
| E⊥c | A | A | A | A |
| E∥c | F | F | F | $A^*$ |
| Silicon vacancy analysed within $C_{3v}$ symmetry: $^4A_2 \leftrightarrow\, ^4A_2$ transition | | | | |
| Polarization | Resonant excitation or ZPL | Phonon assisted (non-resonant) involving phonon of symmetry | | |
| | | $A_1$ | $A_2$ | E |
| E⊥c | F | F | F | A |
| E∥c | A | A | F | F |

We notice that the latter two defects have common symmetry in 4H- and 6H-SiC ($C_{3v}$) and the same structure of the ground and excited states ($^3A_2$ and $^3E$, respectively), hence, their excitation is governed by the same selection rules. On the other hand, $V_{Si}$ has $^4A_2$ ground state in both polytypes. The excited state has two counterparts, $^4A_2$ and $^4E$. In 4H-SiC, the $^4A_2$ is the lowest in energy excited state for both the cubic and hexagonal inequivalent vacancy configurations. The selection rules obtained within the single group $C_{3v}$ for ZPL emission and resonant excitation are those between two $^4A_2$ quartets, yielding emission (or resonant excitation in the ZPL) allowed only with E∥c polarization. Another ZPL termed V1′ appears at higher energy as a companion of the V1 ZPL (associated with $V_{Si}$ at the hexagonal site) at higher temperature, well visible above ~5 K, stemming from the $^4E \leftrightarrow\, ^4A_2$ transition. The selection rules derived within the single group $C_{3v}$ predict E⊥c polarization. We indicate, however, that the selection rules for $V_{Si}$ derived within the single group are only approximate since a proper analysis should be carried out within the double group $\bar{C}_{3v}$ owing to the half-integer spin of the defect.

## IV. RESULTS AND DISCUSSION
### A. Emission selection rules

Although considered previously in different works, we summarize here and display in Table I the polarization selection rules for the ZPLs of the three defects considered here. We notice that the polarization of the ZPLs for the divacancy and the NV pair agree with the group-theoretical analysis (see Table I) and previous publications considering the specific cases of the divacancy in 4H-SiC [20] and the divacancies and NV pairs in 4H- and 6H-SiC [27].

According to the selection rules in $C_{3v}$ symmetry, the polarization of the ZPLs of the axial configurations (PL1 and PL2 in 4H-SiC, QL1, QL2, and QL5 in 6H-SiC) is perpendicular to the axis of the vacancy pair, i.e., to the c-axis ($E \perp c$).

### B. Selection rules with respect to the polarization of non-resonant excitation

We use excitation at 930 nm from a continuous wave Ti-sapphire laser with linearly polarized emission. The polarization of the laser is rotated using λ/2 plate which preserves the linear polarization for any rotation angle. The samples are mounted so that the c-axis is in the focal plane of the objective or the collecting lens, hence the linear polarization of the incident laser can be chosen to conclude any angle with the c-axis. In two cases, the NV pair in 4H-SiC and the divacancy in 6H-SiC, we had bulky samples with parallelepiped shape allowing weakly focused laser beam (by a lens of focal length ~150 mm) on the sample and the use of macroscopic collecting system consisting of a parabolic mirror and a 200-mm focal length lens for focusing on the monochromator slit. Sometimes with bulky samples we have used also rectangular geometry: the **k**-vector of the exciting beam is perpendicular to the c-axis and to the detection direction. Thinner substrates (of thickness 350 – 500 μm) are mounted edge-on and a microscope objective is used to focus the laser on the edge of the sample and collect the emitted PL.

Fig. 1 displays the spectra of the divacancy in 4H-SiC [Fig. 1(a)] and 6H-SiC [Fig. 1(b)] recorded with a polarization of the excitation laser $E_L$ perpendicular and parallel to the c-axis ($E_L \perp c$ and $E_L \| c$, respectively). The most notable feature is the complete vanishing of half of the divacancy lines (in both 4H- and 6H-SiC) when excited with $E_L \| c$ polarization. In complete analogy with the divacancy, half of the ZPLs associated with the NV pair vanish completely when the laser excitation applied to the sample has $E_L \| c$ polarization. This is illustrated in Fig. 2 (part (a) displays the spectra of 4H-SiC, part (b) – in 6H-SiC).

In both polytypes, half of the ZPLs for both considered defects correspond to the number of lines due to the axial configurations. We will show that the vanishing of lines with $E_L \| c$ polarization identifies exactly these configurations and is a consequence of the $C_{3v}$ symmetry and the symmetries of the ground (GS) and the excited (GS) states which are common for these two defects. Moreover, we will demonstrate that photons with $E_L \| c$ polarization are not absorbed at all by the axial configurations, hence neither the ZPLs nor the associated phonon sidebands can appear in the spectra excited with $E_L \| c$.

We refer to Table I, where the selection rules in $C_{3v}$ symmetry for direct and phonon assisted transition between the GS and the ES are listed. Since both defects (divacancy and NV) have integer spin ($S = 1$), the group-theoretical analysis is restricted to the single group $C_{3v}$. We see that the direct transitions for the axial configurations are forbidden with $E \| c$ polarization and allowed with $E \perp c$ polarization, in agreement with theoretical analysis and experimental data for the divacancy in 4H-SiC [20] and our experimental data. Here E denotes the electric field polarization of the ZPLs, hence the same selection rules apply for *resonant* excitation ($E_L \| c$ – forbidden, $E_L \perp c$ – allowed).

Let us consider now the phonon assisted transitions for the axial configurations of the divacancy and the NV pair. There exist three possible symmetries for the phonons in $C_{3v}$, namely, $A_1$, $A_2$, and E. The former two symmetries are one-dimensional, and the atomic displacements are along the c-axis of the crystal, whereas the two-dimensional E representation of $C_{3v}$ describes phonons with displacements in the basal plane, i.e., perpendicular to the c-axis. Table I shows that all phonon-assisted transitions with $E\perp c$ polarization are allowed, whereas for $E\|c$ only transitions assisted with phonons of E symmetry are allowed according to the formal group theoretical analysis. However, light with $E\|c$ polarization cannot interact with phonons of E symmetry because the electric field of the incident photons ($E\|c$) is orthogonal to the atomic displacements of phonons with E symmetry. Consequently, such transitions are physically forbidden.

Thus, we obtain that in the cases of the axial configurations of the divacancy and the NV pair in both 4H- and 6H-SiC, both direct and indirect (phonon assisted) transitions with $E\|c$ polarization are prohibited. When photon absorption is considered, i.e., the polarization E represents the polarization of the incident laser $E_L$, this means that absorption of photons by the axial configurations of the considered defects is prohibited for $E_L\|c$ not only for the direct but also for the indirect (phonon assisted) absorption. This is in complete agreement with the experimental data.

We now discuss the non-axial configurations of the divacancy and NV, a. k. a. basal plane configuration, which have $C_{1h}$ symmetry *per se*. The ground state has A´´ symmetry compatible with the $A_2$ symmetry of the ground state in $C_{3v}$. (Here, we denote the two irreducible representations of $C_{1h}$ with A´ and A´´ to distinguish them from the $A_1$ and $A_2$ representations of $C_{3v}$.) The excited state of E symmetry splits into A´ and A´´ states when the symmetry is lowered to $C_{1h}$. Trivial analysis shows that from a group-theoretical point of view, all transitions are allowed, i.e., with both $E\|c$ and $E\perp c$ polarizations. In addition, for each of the basal configurations, there exist three equivalent orientations of the defect axis (the line connecting the two components of the defect). We notice that if $E\|c$ polarization is considered, the polarization orientation is equivalent for all three orientations of a basal configuration. However, if an arbitrary $E\perp c$ polarization is considered, with photons propagating (nearly) along the c-axis, as is the most common experimental geometry for both the impinging laser excitation and the received PL from the c-face of a sample, the orientation of the polarization is different with respect to the axis of each of the three equivalent orientations. This has some interesting consequences, briefly discussed here in a qualitative manner.

### C. Experimental implications

Let us consider first PL measurements on an ensemble of divacancies, or NV centres. In the usual backscattering-from-surface geometry of a c-plane sample with the c-axis nearly perpendicular to the sample surface, the impinging photons propagate approximately along the c-axis and, therefore, have perpendicular to c polarization ($E_L \perp c$). Whatever the orientation of this polarization with respect to the crystal axis, there will always be three equivalent angles between the axis of the defect and the laser polarization differing by 120°. Thus, if we rotate the laser polarization in the basal plane, we do not expect any polarization dependence, i.e., PL intensity for the basal configuration will have more or less isotropic distribution with the angle of rotation of the polarization.

The situation is drastically different if PL measurements on single defects with basal configuration are considered, as studied in Ref. [22] for single NV pairs. If we rotate the polarization of the incident photons in the basal plane, there will be one direction in which the polarization is nearly parallel to the defect axis. Since the absorption for the axial configurations vanishes when the laser polarization is parallel to the c-axis (i.e., to the defect axis), we may anticipate that also for the basal configurations the absorption will have a minimum when the laser polarization is parallel to the defect axis and maximum when it is perpendicular to it. Hence, we expect a strong dependence of the PL on the polarization orientation of the incident light for basal configurations, which has been observed in [22] for a single configuration of the NV pair in 4H-SiC at room temperature. The single defect with this property has been attributed to an axial configuration (the defect is labelled NV-15 in Ref. [22]), but it likely should be attributed to one of the basal configurations. On the contrary, no strong dependence on the incident light polarization which would involve nearly vanishing of the PL for a certain incident light polarization is expected for the axial configurations (single defects or ensemble), because in the backscattering geometry considered here the laser polarization will be always perpendicular to the c-axis ($E_L \perp c$) and the transition between the ground and excited state is always allowed. The defect labelled NV-1 in [22] has this property, the PL varies by a factor of 2 when the exciting laser polarization is rotated in the basal plane but does not vanish. Hence, this defect should be associated with the axial configural, not with the basal one as suggested in [22].

Using the notion that the axial configurations of VV and NV vanish when excited with $E_L \parallel c$ as a tool for verifying the configuration, we can compare the experimental results presented in Fig. 1 with theoretical calculations. Theoretical data for the ZPLs of the NV centre in 4H-SiC can be found in [15,16,28] and in 6H-SiC in [18]. The theoretical estimates for the ZPL positions of the divacancy are available in [29,30] and [17] for 4H- and 6H-SiC, respectively. The vanishing of the NV2 and NV3 lines when excited with $E_L \parallel c$ implies that these two lines are associated with the axial *hh* and *kk* configurations in 4H-SiC [cf. Fig.1(a)]. The theoretical calculations, however, yield that the lowest energy ZPLs are associated with the axial configurations, indicating that the precision in the ZPL calculations is still insufficient for comparison with experiment [15,28]. Nevertheless, using the D-tensor parameters for identifications of the microscopic NV configurations correctly identifies the ZPL positions of the axial configurations.[16] In [18], the SL2, SL5, and SL6 lines are associated with the axial configurations in 6H-SiC, which is correct according to Fig. 1(b) (notice that Ref. [18] uses different notations for the ZPLs of the NV pair in 6H-SiC). For the divacancy, all theoretical calculations predict that the axial configurations have the lowest energy [29,30]. This is confirmed in Fig. 1(c) since PL1 and PL2 vanish for $E_L \parallel c$ polarization. However, the assignment of the axial configurations for 6H-SiC in Ref. [17] needs adjustment, since the QL5 line identified in this reference as a basal configuration ($hk_1$) does vanish, while the QL6 line suggested to be axial configuration ($k_2k_1$) does not vanish for $E_L \parallel c$ polarization. Since, according to Ref. [17], these two configurations have significant overlap of their Gaussian standard deviations (cf. Fig. 2(b) of Ref. [17]), indicating substantial probability that their ordering in energy is reversed compared to theory, it is most likely that the correct association of QL5 and QL6 is with $k_2k_1$ and $hk_1$ configurations, respectively. Here we use the notations of Refs. [18,31] for the hexagonal (*h*) and cubic ($k_1$, $k_2$) lattice sites in 6H-SiC (Ref. [17] uses somewhat different labelling of the lattice sites, hence $hk_1$ is $hk_2$ and $k_2k_1$ is $k_2k_2$ in Ref. [17]; the order is $V_{Si}V_C$). The corrected assignment for QL5 and QL6 is reflected in Table II.

The results are summarized in Table II for the two defects considered, VV and NV, in 4H- and 6H-SiC. In this table and in Figs. 1 and 2, we use the 'SL' notations of Ref. [27] to denote the NV lines in 6H-SiC because of the following. Firstly, we do not use the 'NV' notations of [18] to avoid confusion with the 'NV' in 4H-SiC and, secondly, in [18] the lines are enumerated in order of ascending wavelength, whereas we prefer to enumerate all the lines for the divacancy and the nitrogen-vacancy pair in order of ascending energies.

Table II. Summary of the line positions and their microscopic identification for the divacancy and the nitrogen-vacancy pair in 4H- and 6H-SiC. The non-resonant (phonon assisted) excitation selection rules stipulate that $E_L \| c$ polarization cannot excite all configurations denoted as axial. The basal configurations can be excited with both $E_L \| c$ and $E_L \perp c$ polarization. The order of the sites is $V_C V_{Si}$ for VV and $N_C V_{Si}$ for NV. The notations for the lattice sites $h$, $k_1$ and $k_2$ for 6H-SiC follow Refs. [18,31] (different from Ref. [17]).

| Line | Position in nm (meV) | Axial/Basal | Identification |
|------|----------------------|-------------|----------------|
| *Divacancy in 4H-SiC* | | | |
| PL1 | 1132.0 (1095.0) | Axial | $hh$ [a] |
| PL2 | 1130.5 (1096.5) | Axial | $kk$ [a] |
| PL3 | 1107.6 (1119.1) | Basal | $kh$ [a] |
| PL4 | 1078.5 (1149.3) | Basal | $hk$ [a] |
| *Divacancy in 6H-SiC* | | | |
| QL1 | 1139.6 (1087.6) | Axial | $k_1 k_2$ [b] |
| QL2 | 1135.0 (1092.1) | Axial | $hh$ [b] |
| QL3 | 1123.9 (1102.9) | Basal | $k_1 h$ [b] |
| QL4 | 1107.4 (1119.3) | Basal | $k_2 k_2$ [b] |
| QL5 | 1093.5 (1133.5) | Axial | $k_2 k_1$ [b,c] |
| QL6 | 1093.0 (1134.0) | Basal | $hk_1$ [b,c] |
| *Nitrogen-vacancy pair (NV$^-$) in 4H-SiC* | | | |
| NV1 | 1242.7 (997.5) | Basal | $hk$ [d] |
| NV2 | 1223.0 (1013.5) | Axial | $kk$ [d] |
| NV3 | 1179.6 (1050.7) | Axial | $hh$ [d] |
| NV4 | 1176.0 (1054.0) | Basal | $kh$ [d] |
| *Nitrogen-vacancy pair (NV$^-$) in 6H-SiC* | | | |
| SL1 | 1240.9 (998.9) | Basal | $hk_1$ [e] |
| SL2 | 1226.4 (1010.7) | Axial | $k_2 k_1$ [e] |
| SL3 | 1203.1 (1030.2) | Basal | $k_2 k_2$ [e] |
| SL4 | 1182.9 (1047.8) | Basal | $k_1 h$ [e] |
| SL5 | 1182.6 (1048.1) | Axial | $hh$ [e] |
| SL6 | 1153.7 (1074.3) | Axial | $k_1 k_2$ [e] |

[a] From Refs. [9,23]. Note that [9] uses the order $V_C V_{Si}$ for labelling.
[b] From Ref. [17].
[c] The QL5 and QL6 lines are reassigned in this work, as explained in text.
[d] According to Ref. [32].
[e] From Ref. [18]

D. Comparison with the silicon vacancy $V_{Si}$

It is instructive to compare the excitation selection rules for the axial divacancy and NV-pair configurations with those for the silicon vacancy $V_{Si}$ which also possesses $C_{3v}$ symmetry but has half-integer spin (S = 3/2). The GS and the ES of the latter defect are both quartets of $A_2$ symmetry, $^4A_2$. The selection rules deduced within the single group $C_{3v}$ stipulate that transitions with $E\perp c$ are forbidden for the ZPLs of both configurations, hexagonal and cubic, while the $E\|c$ transitions are allowed [21,23]. This is also what is approximately observed experimentally (see [33,23]), albeit most experiments can register a weak contribution from the "forbidden" perpendicular to the c-axis polarization. However, the notion that the single-group selection rules are approximately valid also in the description of the silicon vacancy can be misleading, because a proper treatment of the selection rules must be done within the double group $\bar{C}_{3v}$. In fact, we will show below that the observed dominating polarization of the V1 and V2 lines 4H-SiC is most likely associated with the well-known spin polarization of $V_{Si}$ under optical excitation and not with the "approximate validity of the single-group selection rules". In the case of half-integer spin, all substates of the ground and excited states transform as one of the extra representations of the double group, $E_{1/2}$, $^1E_{3/2}$, or $^2E_{3/2}$ [34]. The former two-dimensional representation describes the transformation properties of a wavefunction with spin component $S_z = \pm 1/2$, and the latter two one-dimensional representations are subject to time-reversal (Kramer's) degeneracy and their direct sum $E_{3/2} = {}^1E_{3/2}\oplus{}^2E_{3/2}$ represents the transformation properties of the wavefunctions with $S_z = \pm 3/2$. The selection rules for the ZPL (direct transitions) can be expressed with the following three statements (see, e.g., [35]):

$E_{1/2} \leftrightarrow E_{1/2}$ – allowed with $E\|c$ and $E\perp c$, (1a)

$E_{1/2} \leftrightarrow E_{3/2}$ – allowed with $E\perp c$, and (1b)

$E_{3/2} \leftrightarrow E_{3/2}$ – allowed with $E\|c$. (1c)

Here E denotes the emitted photon polarization if ZPL emission is considered, or $E \equiv E_L$ if resonant absorption is considered. The selection rules for phonon assisted absorption remain unchanged for $E_L\|c$ (only phonons of $A_1$ and $A_2$ symmetry can assist the absorption process). On the other hand, phonon assisted absorption with $E_L\perp c$ can only involve phonons of E symmetry, and in this case all phonon assisted transitions are allowed ($E_{1/2} \leftrightarrow E_{1/2}$, $E_{1/2} \leftrightarrow E_{3/2}$, and $E_{3/2} \leftrightarrow E_{3/2}$). Notice that both quartets, $^4A_2$ (GS) and $^4A_2$ (ES), contain both substates, $S_z = \pm 1/2$ transforming as $E_{1/2}$, and $S_z = \pm 3/2$ transforming as $E_{3/2}$. Since the zero-field splitting (ZFS) between the substates comprising the GS and the ES is in the tens of MHz range, it is unresolvable in optical experiments and in the ZPLs we expect to see contribution from both polarizations, $E\perp c$ and $E\|c$.

Nevertheless, both V1 and V2 manifest dominating $E\|c$ polarization which may be due to the polarization of the ground (and, possibly, excited) states under optical excitation. It has been shown that the $E_{3/2}$ sublevel of the GS has dominant population (above 90 % [36]) under optical excitation [37]. The same dominant population of the $E_{3/2}$ sublevel of the ES can be anticipated, because if the $E_{1/2}$ sublevel of the ES also had substantial population one would expect to see significant contribution in the ZPL from $E\perp c$ polarization from recombination from $E_{1/2}$ to both counterparts $E_{1/2}$ and $E_{3/2}$, in accord with Eqs. (1a) and (1b). Such contribution is present but nearly negligible, compared to the $E\|c$ component of the ZPLs V1 and V2 [33]. Thus, if both the GS and the ES have predominantly radiative recombination with $E\|c$, it is the recombination between the two $E_{3/2}$ counterparts of the ES and GS. Similar hypothesis can be stated for V1´ line. If the ES in this transition has dominant $E_{1/2}$ population, and the recombination to $E_{3/2}$ has larger probability than that to $E_{1/2}$, then we can understand the observed dominant $E\perp c$ polarization in the V1´ line. However, for the moment the only experimental evidence in favour of this hypothesis is that the final state in most of the radiative transitions is $E_{3/2}$, as demonstrated by the spin polarization [36]. Further theoretical work is needed to investigate the probabilities for population of the sublevels in the ES, for both $^4A_2$ and $^4E$, hence the provided possible explanation of the ZPLs' polarizations should be considered hypothetical for the moment.

Experimental data illustrating the presence of E⊥c polarization in the V1 and V2 lines, and weak E∥c component in the V1´ line is given in [33].

### E. Selective excitation of the PL3 line of the divacancy in 4H-SiC

LTPL measurements on an ensemble without special regard for the polarization of the exciting laser usually display all the four ZPL lines PL1 – PL4 if the laser energy is higher than the energy of PL4 (the highest-energy ZPL). As discussed earlier, if excitation with $E_L$∥c polarization is used the axial configurations with ZPLs PL1 and PL2 together with their pertaining phonon side bands (PSB) are not excited at all. If, in addition, the laser energy is chosen between the energy positions of PL3 and PL4, then only the PL3 line will be excited. Thus, a single ZPL together with its PSB for the divacancy in 4H-SiC can be isolated in the PL spectrum of divacancies ensemble containing all configurations. The situation is like that of observation of the PL from single defect (in this case, PL3 corresponding to the *hk* configuration of $V_{Si}V_C$ in 4H-SiC), but with the benefit of the much stronger signal associated with the ensemble.

This selective excitation of the PL3 line in an ensemble is depicted in Fig. 3(a). The upper spectrum is obtained with excitation at 1090 nm (1.1371 eV) in between the PL4 (1.1493 eV) and PL3 (1.1191 eV) lines. The laser is filtered using two sharp edge long pass filters with a cutoff at 1100 nm. The bottom spectrum obtained with a common excitation of 930 nm (1.333 eV) above the energy of all four ZPLs is given for comparison. Fig. 3(a) shows two ODMR spectra obtained with the same microwave antenna configuration but different excitation wavelengths, non-selective excitation with laser at 1020 nm (lighter thicker curve) and selective excitation with 1090 nm (darker thinner curve). In both cases $E_L$∥c is used, so that the former ODMR spectrum contains contribution from only PL3 and PL4 (PL1 and PL2 are not excited), while the selectively excited spectrum contains the pure PL3 contribution displayed in the top spectrum of Fig. 3(a). We point out two features of the experiment with selective excitation of PL3. Firstly, the ODMR contrast is improved by at least a factor of two as seen from the inset (Fig. 3), which is anticipated because the background contribution from PL4 and its phonon sideband to the ODMR signal of PL3 is removed. The use of excitation with $E_L$ not parallel to the c-axis will further deteriorate the contrast due to added contribution from PL1 and PL2 to the background. The second feature concerns the Debye-Waller (DW) factor of the PL3 line which shows a significant increase compared to the spectrum obtained with common excitation at 930 nm. The origin of this improvement is not understood at present and a thorough investigation of the factors influencing the DW factor is beyond the scope of this work. Nevertheless, the possibility of dealing with a single selectively excited ZPL with intensity typical for an ensemble may prove beneficial in future sensor applications.

### V. SUMMARY

In conclusion, we have demonstrated theoretically and experimentally that the high-symmetry (axial) configurations of the divacancy and the NV pair in both 4H and 6H SiC are highly sensitive to the polarization of the exciting laser. No absorption and hence no excitation of the axial configurations occurs if the polarization of the excitation $E_L \parallel c$. This is shown to be due to the fact that only phonons of E symmetry may participate in allowed transitions between the ground and excited states for $E_L \parallel c$, according to the group theoretical selection rules. However, phonons with in-plane displacements cannot interact with photons of $E_L \parallel c$ polarization since the electric field of the electromagnetic wave is orthogonal to the displacement of phonons of E symmetry. A comparison with the properties of the silicon vacancy $V_{Si}$ which also has GS of orbital $A_2$ symmetry and ES of E symmetry for some configurations, but possesses half-integer spin, shows that excitation of the corresponding configurations is not prohibited. This is attributed to the necessity of analysing the selection rules within the double group $\bar{C}_{3v}$. The energy ordering of the divacancy ZPLs in 4H-SiC allows selective excitation of only one of the four configurations (PL3 line) using suitable laser energy and $E_L \parallel c$ polarization which can be useful for using the PL3 line in sensor applications with an ensemble of divacancies.

**Acknowledgements.** Financial support from the Knut and Alice Wallenberg Foundation (KAW 2018.0071) and the EU H2020 Project QuanTELCO (No. 862721) is acknowledged. D.S. acknowledges support from the AFM (CeNano grant 2021). J.D. acknowledges support from the Swedish e-science Research Centre (SeRC) and the Swedish Research Council (VR) Grant No. 2022-00276. T.O. thanks for the support from the Japan Society for the Promotion of Science JSPS KAKENHI (grant Nos. 20H00355 and 21H04553).

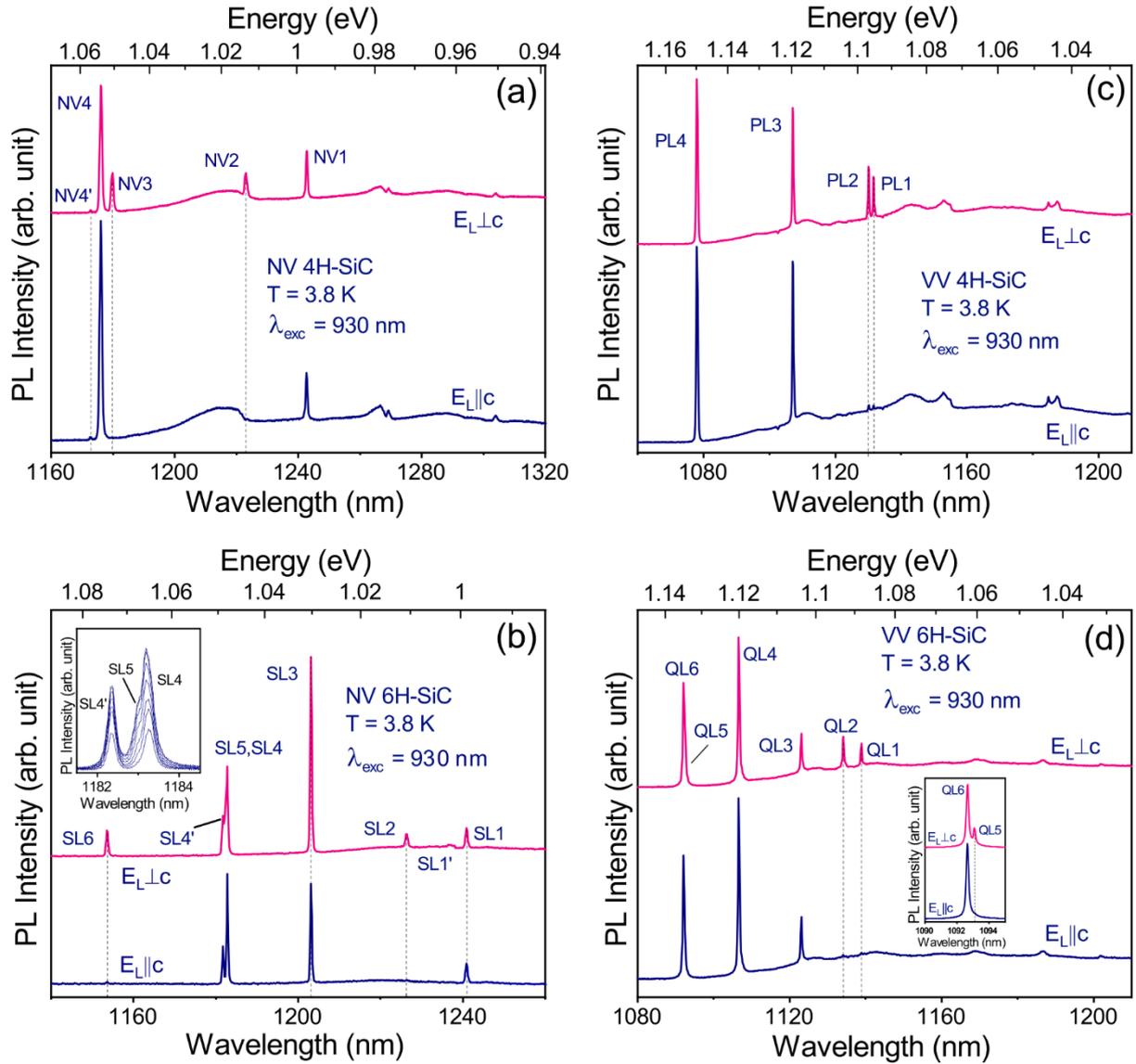

Fig. 1. Low-temperature PL spectra of the NV pair [panels (a), (b)] and the divacancy [panels (c), (d)] in 4H-SiC [panels (a), (c)] and 6H-SiC [panels (b), (d)]. The ZPLs due to the axial configurations with $C_{3v}$ symmetry all vanish for polarization of the excitation $E_L \parallel c$. The insets in panels (b) and (d) show spectra recorded with higher resolution; the latter zooms on the nearly degenerated QL5 and QL6 lines, while the former shows the angular dependence of the SL4 and SL5 lines recorded with 15° steps between $E_L \perp c$ (topmost curve) and $E_L \parallel c$ (bottom curve). NV4´, SL1´, and SL4´ denote PL lines originating from closely spaced counterparts of the excited state detectable at temperatures ~ 4 K. For the NV ZPLs in 6H-SiC we use the 'SL' notations of Ref. [27] to avoid confusion with the NV lines in 4H-SiC.

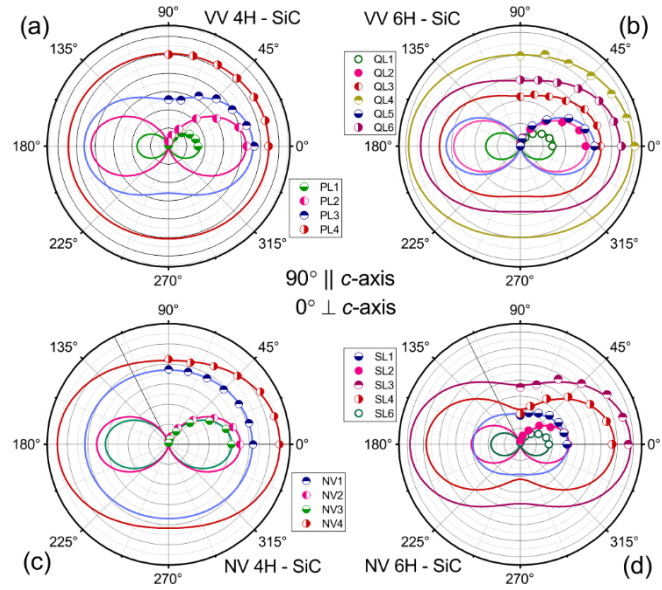

Fig. 2. Angular dependence of the ZPL intensity on the angle $\varphi$ between the laser polarization and the c-axis for the divacancy and the nitrogen vacancy pair in 4H and 6H polytypes, as denoted for each polar plot. The experimental data is fitted with an expression of the form $I(\varphi) = A(1 + B \cos 2\varphi)$, where $I$ is the ZPL intensity and $A$ and $B$ are fitting constants specific for each ZPL. $B = 1$ for the axial configurations. The laser polarization is $E_L \perp c$ at $\varphi = 0°$ and $E_L \| c$ at $\varphi = 90°$.

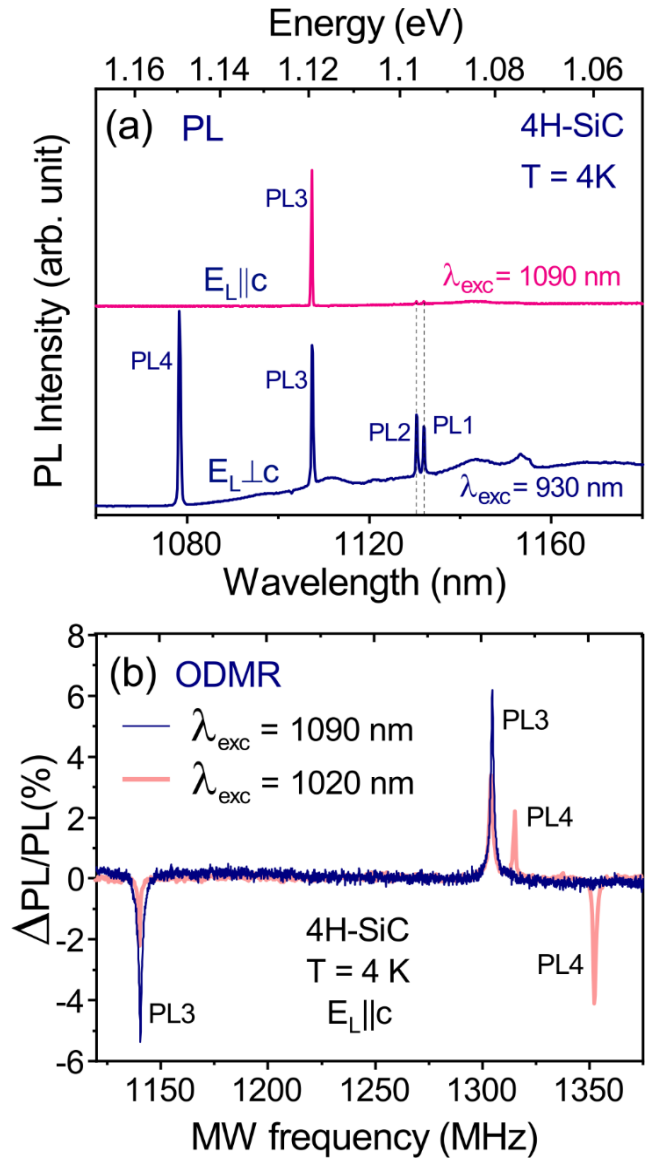

Fig. 3. Selective excitation of the PL3 line in ensemble of divacancies in 4H-SiC. (a) Selectively excited spectrum of PL3 line (top curve) using excitation with $E_L \parallel c$ polarization and energy lower than the PL4 line (laser at 1090 nm, or 1.137 eV). PL1 and PL2 are not excited because there is no absorption for these configurations with $E_L \parallel c$ polarization. The bottom spectrum showing all four divacancy lines and their phonon sidebands is provided for comparison (excitation with $E_L \perp c$ at 930 nm). Note the enhanced Debye-Waller factor with selective excitation of PL3. (b) ODMR spectra obtained using the selective excitation of PL3 only (dark thinner curve) and of PL3 and PL4 using higher laser energy (light thicker curve), illustrating the contrast improvement for PL3 when the background emission of PL4 is removed, as discussed in text.

# Selection rules in the excitation of the divacancy and the nitrogen-vacancy pair in 4H- and 6H-SiC

Danial Shafizadeh[1], Joel Davidsson [1], Takeshi Ohshima[2,3], Igor A. Abrikosov[1], Nguyen T. Son[1], and Ivan G. Ivanov[1]

[1)]Department of Physics, Chemistry and Biology, Linköping University, SE-58183, Linköping, Sweden

[2)] National Institutes for Quantum Science and Technology, 1233 Watanuki, Takasaki, Gunma 370-1292, Japan

[3)] Department of Materials Science, Tohoku University, 6-6-02 Aramaki-Aza, Aoba-ku, Sendai 980-8579, Japan

1. Selection rules in $C_{3v}$ symmetry

We use the well-known group-theoretical result that the matrix element describing dipole transition between two discrete states with wavefunctions $\Psi_i$ and $\Psi_f$ can be nonzero only if the direct product $\Gamma_f \otimes \Gamma_d \otimes \Gamma_i$ contains the trivial representation $\Gamma_1$:

$$\Gamma_f \otimes \Gamma_d \otimes \Gamma_i \ni \Gamma_1. \tag{S1}$$

Here $\Gamma_f$, $\Gamma_i$, and $\Gamma_d$ denote the representations according to which the wavefunctions of the final and the initial state, and the dipole operator are transforming. $\Gamma_1$ (or $A_1$) is the trivial representation. The dipole operator transforms as the electric field vector $E_L$ of the exciting electromagnetic field. Thus, when the exciting polarization is parallel to the c-axis ($E_L \| c$), $\Gamma_d$ is $A_1$, and when $E_L \perp c$ $\Gamma_d$ is the two-dimensional E representation in $C_{3v}$ symmetry.

For the divacancy and the NV pair the ground state of the direct transition $\Psi_i$ is triplet $^3A_2$ with $A_2$ symmetry and the excited state $\Psi_f$ is triplet $^3E$ with E symmetry. Thus, we obtain the selection rules for direct transitions $^3A_2 \leftrightarrow {}^3E$:

For $E_L \perp c$: $E \otimes E \otimes A_2 = A_1 \oplus A_2 \oplus E \ni A_1$ – allowed. (S2)

For $E_L \| c$: $E \otimes A_1 \otimes A_2 = E \neq A_1$ – forbidden. (S3)

Considering phonon-assisted transitions (absorption or emission), the representation according to which the assisting phonon displacements are transforming adds as another direct product to the left side of Eqs. (S2 – S3). The phonon displacements transform as either $A_1$, $E_2$, or E in $C_{3v}$. For instance, if we check if phonons of $A_2$ symmetry can assist absorption or emission with $E_L \| c$ we examine the product $A_2 \otimes E \otimes A_1 \otimes A_2 = E$ and the process is forbidden. The final result for phonon assisted transitions between $^3A_2 \leftrightarrow {}^3E$ is given in Table I of the main article. We notice that there is only one entry in this Table for which the transition with $E_L \| c$ assisted with phonons of E symmetry is formally allowed by group theory, but physically unfeasible because the phonon displacements are orthogonal to the electric-field vector of the light, as discussed in main text. In all other cases the allowed phonon-assisted transitions involve phonons with displacements collinear with the electric field vector of the light. Finally, we notice that we do not specify whether the phonons involved are lattice phonons or local phonons; the selection rules are valid in both cases.

## 2. Selection rules for the silicon vacancy ($V_{Si}$)

As discussed in the main text, the selection rules for $V_{Si}$ are subject to derivation within the double group $\bar{C}_{3v}$ due to the half-integer spin of the electron ($S = 3/2$). In the double group, each of the ground state and the excited state has two counterparts of symmetries $E_{1/2}$ and $E_{3/2}$, where $E_{1/2}$, $^1E_{3/2}$ and $^2E_{3/2}$ ($E_{3/2} = {}^1E_{3/2} \oplus {}^2E_{3/2}$) are the extra representation of $\bar{C}_{3v}$. Consequently, all transitions become allowed, $^4A_2 \leftrightarrow {}^4A_2$, and $^4A_2 \leftrightarrow {}^4E$, valid for both direct and phonon assisted transitions.

The ZPLs stemming from the transition $^4A_2 \to {}^4A_2$ are polarized predominantly E∥c, and the ZPLs due to the transition $E \to {}^4A_2$ have dominant E⊥c polarization, albeit weak contribution from the complementary polarization can also be discerned. This is illustrated in Fig. S1 for the case of $V_{Si}$ in 4H-SiC. The case of $V_{Si}$ in 6H-SiC is considered in greater detail in [S1] where it is also demonstrated that the contribution from the "forbidden" by the single-group analysis is weak, but non-negligible.

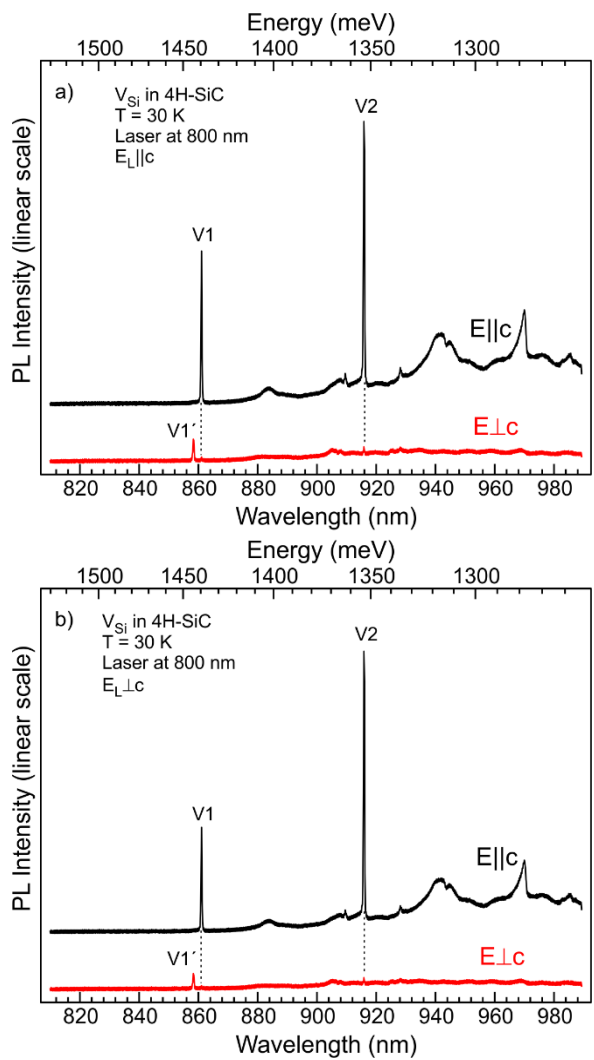

Fig. S1. Polarized spectra of the silicon vacancy ($V_{Si}$) in 4H-SiC obtained with two polarizations of the exciting laser, (a) $E_L$∥c, and (b) $E_L$⊥c. The contributions from the E⊥c polarization in the V1 and V2 lines, albeit weak, are distinguishable.